# Revisiting Phase Diagrams of Two-Mode Phase-Field Crystal Models

Arezoo Emdadi[1], Mohsen Asle Zaeem[1*] and Ebrahim Asadi[2]

[1]Department of Materials Science and Engineering, Missouri University of Science and Technology, Rolla, MO 65409, USA
[2]Department of Mechanical Engineering, University of Memphis, Memphis, TN 38152, USA

**Abstract**: In this work, phase diagrams of a modified two-mode phase-field crystal (PFC) that show two-dimensional (2D) and three-dimensional (3D) crystallographic structures were determined by utilizing a free energy minimization method. In this study the modified two-mode PFC model (presented by E. Asadi and M. Asle Zaeem, Comput. Mater. Sci. 2015) was used, in which the free energy can be exactly minimized in each stable crystal structure allowing calculation of accurate phase diagrams for two-mode PFC models. Different crystal structures, such as square, triangle, body-centered cubic (bcc), face-centered cubic (fcc), and stripe lattice structures as well as their coexistence regions were considered in the calculations. The model parameters were discussed to calculate phase diagrams that can be used as a guideline by other researchers for studying solidification and solid state phase transformation using two-mode PFC model.

Corresponding author:  zaeem@mst.edu (M. Asle Zaeem)

## 1. Introduction

The phase-field crystal (PFC) model is a reformulation of the Swift-Hohenberg [1] Equation, a model for simulation of non-conserved thermal fluctuation fields in the Rayleigh-Benard convection problem [2], with conserved dynamics. The PFC model contains atomistic scale details and works on diffusive time scales, and it can be directly derived by approximations from density functional theory [3]. Therefore essential physics of the material such as elasticity, plasticity, dislocation and grain boundary formation are inherently incorporated in the PFC model. PFC models were successfully utilized in many different studies in materials science [3]. Different phenomena such as solidification [4, 5], binary alloy crystallization [6-9], Kirkendall effect [10], and grain-boundary premelting [11] were studied by utilizing different PFC models. Also different properties such as the bulk modulus and grain-boundary energies [12], crystal-melt interfacial free-energy [12-14], and stacking faults [15] were calculated using this robust model.

The original PFC model is frequently called one-mode PFC model because its free energy functional damps the dynamics of the system except near the first density wave vector [16, 17]. The free energy functional of the one-mode PFC model is:

$$F = \int \left\{ \frac{1}{2} \phi \left[ \alpha + \lambda (q_0^2 + \nabla^2)^2 \right] \phi + \frac{g}{4} \phi^4 \right\} d\boldsymbol{r} \,, \tag{1}$$

where $\phi$ is a function of spatial positions related to the density field, $q_0$ is the magnitude of the principal reciprocal lattice vectors (RLVs) and $\alpha$, $\lambda$ and $g$ are model parameters. It is convenient to use the dimensionless form of the free energy by these relationships: $\varepsilon = -\alpha / \lambda q_0^4$, $\psi = \phi \sqrt{g / \lambda q_0^4}$, $\boldsymbol{x} = q_0 \boldsymbol{r}$ and $F^* = \frac{g}{\lambda^2 q_0^5} F$, which result in the non-dimensional free energy functional of the one-mode PFC model:

$$F^* = \int \left\{ \frac{1}{2} \psi \left[ -\varepsilon + (1 + \nabla^2)^2 \right] \psi + \frac{\psi^4}{4} \right\} d\boldsymbol{x} \,, \tag{2}$$

where $\varepsilon$ is a small parameter, and $\psi$ is the dimensionless density field which is the summation of the average density of the solid state, $\bar{\psi}_s$, and a periodic function representing density fluctuations around $\bar{\psi}_s$. The dimensionless density field for different crystal structures will be explained in Section 2 and Appendix A. Fig. 1 shows the two-dimensional (2D) and three-dimensional (3D) phase diagrams of the one-mode PFC model using the conventional free energy functional in Eq. (1) [12, 17]. The phase diagrams of PFC models show that which crystal structures are stable at different values of the model parameter, $\varepsilon$. This type of PFC can present hexagonal close packed (hcp) or triangle, body-centered cubic (bcc), and strip structures but it can not present face-centered cubic (fcc) and square crystal structures. Jaatinen et al. [18] added a cubic term ($-b\phi^3 / 3$, where $b$ is another model parameter) to Eq. (1) and showed that with a particular choice of the modified $\varepsilon (= (b^2 / 3g - \alpha) / \lambda q_0^4)$ in Eq. (2), hcp, bcc and fcc lattice structures can be stable in the one-mode PFC, but no coexistence between fcc and bcc can be achieved, also square crystal structure is not stable in Jaatinen et al. [18] model.

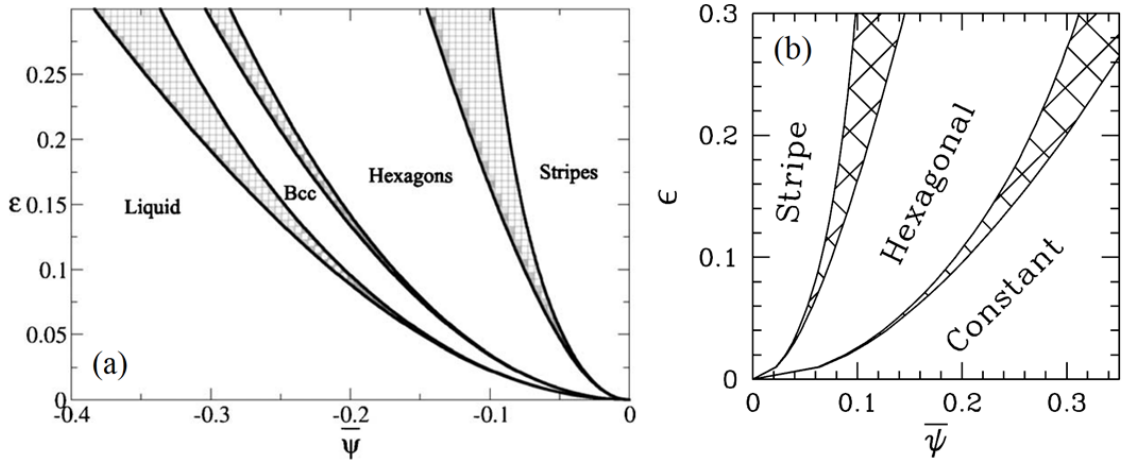

**Fig. 1.** Phase diagram of the one-mode PFC model in (a) 3D [12] and (b) 2D [17]; Constant region is liquid phase.

Even though one-mode PFC models predict some properties such as solid-liquid interface free energy and grain boundary free energy with a good agreement with experimental data, they have difficulties in accurately predicting some other properties such as expansion in melting [19]. To improve the accuracy of the results of quantitative PFC, the effect of second or higher density wave vectors in the free energy functional needs to be considered. In two-mode PFC, as proposed by Wu et al. [20], a second density wave vector was considered. The free energy functionals of the two-mode PFC in dimensional and dimensionless forms are respectively:

$$F = \int \left\{ \frac{1}{2} \phi \left[ \alpha + \lambda (q_0^2 + \nabla^2)^2 \left[ (q_1^2 + \nabla^2)^2 + r_1 \right] \right] \phi + \frac{g}{4} \phi^4 \right\} d\mathbf{r} \,, \tag{3a}$$

$$F^* = \int \left\{ \frac{\psi}{2} \left[ -\varepsilon + (1 + \nabla^2)^2 \left[ (Q_1^2 + \nabla^2)^2 + R_1 \right] \right] \psi + \frac{\psi^4}{4} \right\} d\mathbf{x} \,, \tag{3b}$$

where $R_1 = r_1 / q_0^4$ and $Q_1 = q_1 / q_0$. In Eq. (3a), $q_0$ corresponds to the principal RLVs of the crystal structure and $q_1$ to some other set of RLVs with larger wave vector magnitude. $r_1$ is the model parameter which can be positive or negative to provide flexibility to get stability of different crystal structures. For example for fcc, the principal RLVs is related to [111] and the second one is [200], so $Q_1 = \sqrt{4/3}$. Wu et al. [20] two-mode PFC model has two degree of freedoms (DOFs) or independent model parameters, $\varepsilon$ and $R_1$.

For $R_1 = 0$, Eq. (3b) reduces to the free energy functional of Lifshitz and Petrich [21] and 3D phase diagram of the model only exhibits fcc-liquid coexistence. The calculated 3D phase diagram for this model is presented in Fig. 2.

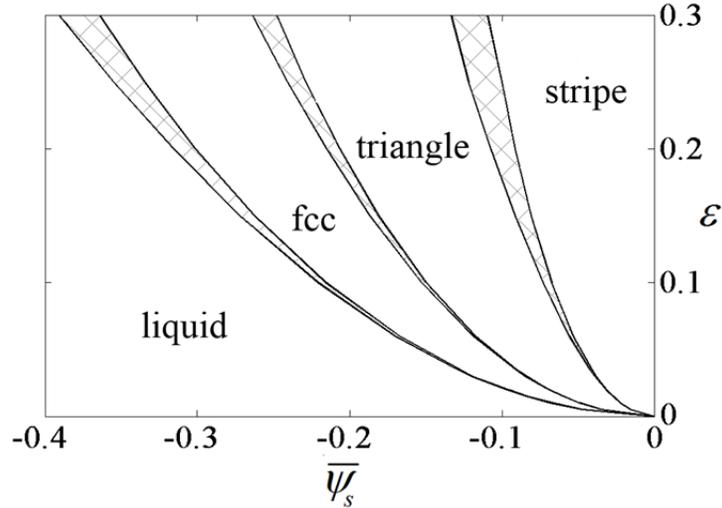

**Fig. 2.** Calculated phase diagram for two-mode PFC with $R_1 = 0$ in Eq. (3b).

By increasing $R_1$ in Eq. (3b), the amplitude of the second mode decreases and the two-mode PFC model reduces to the one-mode PFC model. Wu et al. [20] showed that a small finite value of $R_1$ is required for the phase diagram (for the case of $Q_1 = \sqrt{4/3}$) to have both bcc-liquid and fcc-liquid coexistence, and also to make this model capable of studying phase transformation from bcc to fcc (and vice versa). The computed phase diagram including bcc and fcc with $R_1 = 0.05$ for this model is presented in Fig. 3(a) [20], however, this phase diagram was computed based on some assumptions, which will be discussed in Section 2. Wu et al. [20] mentioned that a square crystal structure can be stable in the 2D version of their model, but they did not calculate the 2D phase diagram of their model to show for which range of model parameters this crystalline can be stable. In the following section, we will explain why it is necessary to revisit this phase diagram, and then we will study the effects of different model parameters on 2D and 3D phase diagrams of two-mode PFC.

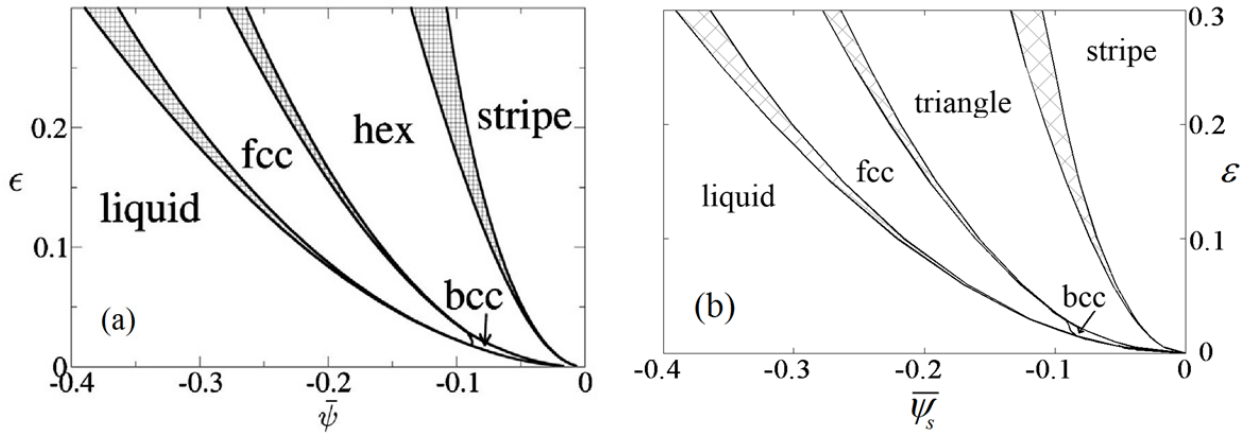

**Fig. 3.** Calculated phase diagram for two-mode PFC (a) from Ref. [20] and (b) for the modified two-mode PFC model with $R_1 = 0.05$ and $R_0 = 0$ [22].

Mkhonta et al. [23] introduced a multimode PFC model, and they showed a system including three length scales can order into five Bravias lattices and other structures such as honeycomb, and kagome. The dimensionless free energy functional and the dynamic equation on diffusive time scales are respectively:

$$F = \int d\vec{r} \left\{ \frac{\psi}{2} \left( r + \lambda \prod_{i=0}^{N-1} \left[ \left( Q_i^2 + \nabla^2 \right)^2 + b_i \right] \right) \psi - \frac{\tau}{3} \psi^3 + \frac{\psi^4}{4} \right\},$$ (4a)

$$\partial \psi / \partial t = \nabla^2 \left\{ \left( r + \lambda \prod_{i=0}^{N-1} \left[ \left( Q_i^2 + \nabla^2 \right)^2 + b_i \right] \right) \psi - \tau \psi^2 + \psi^3 \right\},$$ (4b)

Where $r$, $\lambda$, $b_i$ and $\tau$ are phenomenological constants and $Q_i$ are the magnitude of wave vectors. Mkhonta et al. [23] examined 2D nonequilibrium phase transitions with $N = 3$ (three-mode PFC) by solving the PFC evolution equation numerically. In the case of $N = 2$, this PFC model considers the first two wavelength vectors similar to the two-mode PFC model of Wu et al. [20], but it has five DOFs ($r$, $\lambda$, $b_0$, $b_1$ and $\tau$). The three additional DOFs increase the computational expenses exponentially.

In a recent work, the modified two-mode PFC model (M2PFC) was introduced by E. Asadi and M. Asle Zaeem [22], which has the same DOFs as Wu et al. two-mode PFC model [20], but it has a dependent parameter that enable exact minimization of the free energy in different phases. This model and the method to determine its phase diagrams will be explained in the next section.

## 2. Phase Diagram Calculations

In density functional theory (DFT), the density of the crystalline state can be expressed in terms of the reciprocal lattice vectors (RLVs), $\vec{k}$, and their amplitudes, $A_{\vec{k}}$, by

$$\psi = \bar{\psi}_s + \sum_{\vec{k}} A_{\vec{k}} e^{i\vec{k}\cdot\vec{r}} + c.c. ,$$ (5)

In Eq. (5), $\bar{\psi}_s$ is the average density in the solid state, $\vec{r}$ is the position vector, $A_{\vec{k}}$ is the Fourier amplitudes of the related RLVs, $i = \sqrt{-1}$, and c.c. means complex conjugate. The dimensionless density profiles in the solid state for square, triangular, stripe, fcc and bcc lattice structures in PFC model by considering the first and second wavelength amplitudes are:

$$\psi_{sq} = \bar{\psi}_s + 2A_s(\cos qx + \cos qy) + 4B_s(\cos qx \cos qy),$$ (6a)

$$\psi_{tri} = \bar{\psi}_s + A_s(\cos qx \cos \frac{qy}{\sqrt{3}} - \frac{1}{2}\cos \frac{2qy}{\sqrt{3}}) + B_s(\cos qx \cos \sqrt{3}qy - \frac{1}{2}\cos 2qx),$$ (6b)

$$\psi_{str} = \bar{\psi}_s + A_s \cos qx + B_s \cos 2qx,$$ (6c)

$$\psi_{fcc} = \bar{\psi}_s + 8A_s \cos qx \cos qy \cos qz + 2B_s(\cos 2qx + \cos 2qy + \cos 2qz),$$ (6d)

$$\psi_{bcc} = \bar{\psi}_s + 4A_s(\cos qx \cos qy + \cos qx \cos qz + \cos qy \cos qz)$$
$$+ 2B_s(\cos 2qx + \cos 2qy + \cos 2qz),$$ (6e)

where $A_s$ and $B_s$ are density amplitudes for the first and second RLVs, and $q$ depends on the crystal structure (Appendix A). For square and stripe lattice structures, $q = 1$, for the triangular, $q = \sqrt{3}/2$, for bcc, $q = 1/\sqrt{2}$ and for fcc, $q = 1/\sqrt{3}$. It is worth mentioning that $A_s$ and $B_s$ are different for different crystal structures in Eqs. (6a)-(6e), and they will be calculated by the minimizing free energy density with respect to $A_s$ and $B_s$. The above form of dimensionless density fields is known as the two-mode expansion of density; if only the first wavelength amplitude was considered, in other words $B_s = 0$, then it becomes the one-mode expansion of density. The procedure for deriving Eqs. (6a)-(6e) is explained in Appendix A.

In the phase diagram calculated by Wu et al. [20], two-mode expansion of the density field was only considered for fcc crystal structure, and one-mode expansion of the density field was considered for the other crystal structures. With these assumptions they showed that their model could predict the coexistence of bcc and fcc structures.

The free energy density for every crystal structure can be calculated by substituting the appropriate density profile from Eqs. (6a)-(6e) in Eq. (4a), integrating over the crystal unit cell, and then dividing the resultant by the unit area (in 2D) or unit volume (in 3D). For example the calculated free energy density for fcc crystal structure is:

$$f_{fcc} = -\left[\varepsilon - (\frac{16}{9} + R_1)\right]\frac{\bar{\psi}_s^2}{2} + 4(-\varepsilon + 3\bar{\psi}_s^2)A_s^2 + 3(3\bar{\psi}_s^2 + \frac{1}{9}R_1 - \varepsilon)B_s^2 + \frac{\bar{\psi}_s^4}{4} + 72A_s^2 B_s \bar{\psi}_s$$
$$+ 144A_s^2 B_s^2 + 54A_s^4 + \frac{45}{2}B_s^4.$$ (7)

The dimensionless free energy densities for other crystal structures can be calculated from equations in Appendix B. For fcc crystal structure, the free energy density in Eq. (7) needs to be minimized with respect to the crystal structure. Minimization of the free energy density with respect to density wavelength amplitudes ($A_s$ and $B_s$) gives the relationship of these unknowns in terms of average density of the solid state ($\bar{\psi}_s$) and the model parameters $R_1$ and $\varepsilon$. If two-mode expansion of dimensionless density was considered, the total free energy density would not be minimized with respect to $q$, unless $R_1 = 0$ [22]. For example for fcc crystal structure with free energy density of Eq. (7), $df_{fcc}/dq\,|_{q=1/\sqrt{3}} = 16B_s^2 R_1/\sqrt{3}$. This error increases by increasing $R_1$ or $B_s$. Thus for every crystal structure with two-mode expansion of density, its relevant $q$ will not minimize the free energy density accurately. E. Asadi and M. Asle Zaeem [22] proposed a modified two-mode PFC model (M2PFC) which incorporates two independent parameters, $\varepsilon$ and $R_1$, and one dependent parameter, $R_0$, to prevent this error in minimization of the free energy density in solid crystalline. The dimensional and dimensionless free energy functionals of M2PFC are:

$$F = \int \left\{ \frac{1}{2} \phi \left\{ \alpha + \lambda \left[ (q_0^2 + \nabla^2)^2 + r_0 \right]\left[ (q_1^2 + \nabla^2)^2 + r_1 \right] \right\} \phi + \frac{g}{4} \phi^4 \right\} d\boldsymbol{r} \,, \tag{8a}$$

$$F^* = \int \left\{ \frac{\psi}{2} \left\{ -\varepsilon + \left[ (1 + \nabla^2)^2 + R_0 \right]\left[ (Q_1^2 + \nabla^2)^2 + R_1 \right] \right\} \psi + \frac{\psi^4}{4} \right\} d\boldsymbol{x} \,, \tag{8b}$$

where $R_0 = r_0/q_0^4$, which can be calculated by minimizing the free energy density of the solid state for every crystal structure with respect to $q$ by considering its relevant $q$. $R_0$ is not an independent parameter and can be calculated according to the crystal structure and other model parameters such that the free energy is minimized accurately. For example, for fcc crystal structure $R_0 = 2R_1(B_s^2/A_s^2)$. M2PFC model will be reduced to the Wu et al. [20] two-mode PFC model when $R_0 = 0$. Although Wu et al. model [20] added a new parameter ($R_1$, which is an extra degree of freedom) to Swift-Hohenberg two-mode PFC model [1] to produce fcc crystals and achieve coexistence of fcc and bcc crystallines, this model did not exactly minimize the free energy in fcc crystals when $R_1$ was not zero. The M2PFC model does not add an extra degree of freedom to Wu et al. model [20], the new parameter ($R_0$) is a function of other parameters ($R_1$, etc.), and M2PFC produces stable phases by exactly minimizing the free energy in each phase. The Mkhonta's model [23] with $N=2$ (i.e. two-mode) seems to have similar equations to M2PFC model, but in this model $b_0$ and $b_1$ (which are similar to $R_0$ and $R_1$ in M2PFC model) are independent parameters; although Mkhonta's model [23] has the same wave numbers as M2PFC model, it has three more degree of freedoms. The M2PFC with the same degree of freedom as the Wu et al. [20] model, produces stable phases, all its parameters are connected to physical quantities [19], and it was showed that this model is more accurate quantitatively than previous results [19]. Therefore M2PFC not only gives stable fcc and co-existence between bcc and fcc, but also it gives better quantitative results. Fig. (3b) shows the calculated phase diagram of M2PFC with the assumptions in Ref. [20] and for $R_1=0.05$ and $R_0=0$; this phase diagram is the same as Fig. (3a).

To calculate the phase diagrams, a positive, negative or zero value for $R_1$ is considered, and $\varepsilon$ is changed from 0 to 0.5. Then, for any average densities in solid ($\bar{\psi}_s$) and liquid ($\psi_l$) states, the dimensionless free energy density for each crystal structure and liquid are calculated using Eqs. (B1)-(B6) in Appendix B. In 3D phase diagrams, we chose $R_1=0.05$ as a positive value of $R_1$, because then our results can be compared to the results of Ref. [20]. For a negative value of $R_1$, we chose $R_1=-0.015$, because for larger negative values of $R_1$ the numerical calculations of free energy minimization do not converge for small values of $\varepsilon$. To calculate the 2D phase diagrams, $R_1=\pm0.05$ and $R_1=\pm0.15$ are chosen. The density amplitudes ($A_s$ and $B_s$) can be determined by minimizing the free energy density. The coexistence of solid-liquid was numerically calculated by using the standard common tangent construction [3] by equating the chemical potentials $f_s^{'}(\bar{\psi}_s)=f_l^{'}(\psi_l)=\mu_E$ and grand potentials $f_s(\bar{\psi}_s)-\mu_E\bar{\psi}_s=f_l(\psi_l)-\mu_E\psi_l$ for any phases. Similar procedure was followed to calculate the coexistence regions between two solid phases.

## 3. Results and Discussions

### 3.1. Phase diagram in 3D

I.  Phase diagram of M2PFC model considering two-mode expansion of density field for fcc and one-mode expansion of density for bcc, triangle and stripe

    In Eq. (8b), the total free energy of the solid states is affected by parameter $Q_1$. This parameter is different for different crystal structures, thus it can significantly change the total free energy. In this section to be able to include different crystal structures in one phase diagram, $Q_1$ for one crystal structure is used to determine the parameters for all the other crystal structures (first assumption of Wu et al. [20]). However, to determine the coexistence of the solid-liquid for any crystal structure, it is necessary to consider $Q_1$ for that crystal structure and use $R_0$ according to Eqs. (9a)-(9d). To calculate the phase diagram of M2PFC, $Q_1$ of fcc crystal is considered as the reference, and the values of $R_0$ were calculated for the other crystal structures. Even by considering another $Q_1$ for a crystal structure rather than its own $Q_1$, M2PFC has the ability to choose $R_1$ in a way that the exact minimum free energy for that crystalline is achieved.

    As it was mentioned previously, in M2PFC model, it is necessary to calculate the value of $R_0$ for each crystal structure to minimize the free energy density exactly. This parameter is equal to zero if we consider one-mode expansion of the density. So if we want to recalculate the phase diagram of Fig. 3 considering M2PFC model, and since two-mode expansion density was used only for fcc crystal in Fig. 3 (second assumption of Wu et al. [20]), we need to only calculate $R_0$ for fcc crystal structure. Figs. 4(a-c) show the calculated phase diagrams of M2PFC for $R_1 = 0.05$, $R_1 = 0.1$ and $R_1 = -0.015$ using two-mode expansions of the crystal density field for fcc and one-mode for bcc, triangle and stripe. As it can be seen, this exact minimization shows that, for example for even a small value of $R_1 = 0.05$, the region for stable bcc is smaller than that of the Wu et al. two-mode PFC model [20], Fig. 3(a). In quantitative PFC simulations, the results can be significantly affected by this small changes in the phase diagram, especially noticing that the most of the quantitative PFC simulations have been done so far for small $\varepsilon$ [12-14, 19]. As $R_1$ increases, the region for bcc will increase and have more coexistence region with

the fcc crystal structure. For the negative value of $R_1$, the effect of second wavelength vector increases, and bcc crystal structure is not stable.

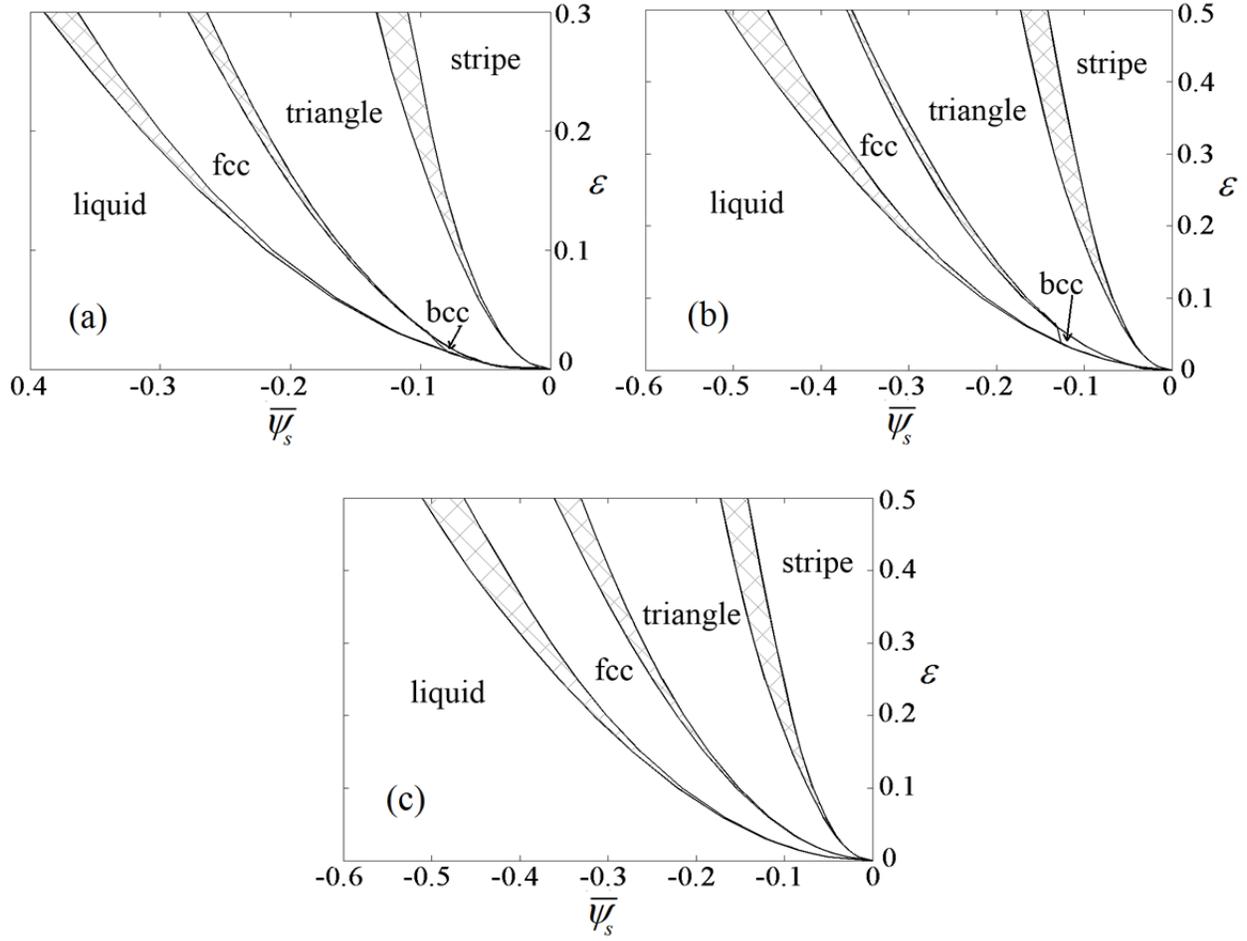

**Fig. 4.** Phase diagrams of the M2PFC model computed using two-mode expansions of the crystal density field for fcc and one-mode for bcc, triangle and stripe; $Q_1 = Q_{1,fcc} = \sqrt{4/3}$; (a) $R_1 = 0.05$, (b) $R_1 = 0.1$, and (c) $R_1 = -0.015$.

Fig. 5 shows phase diagrams of M2PFC by considering $Q_1$ for bcc crystal structure. In this case fcc crystal structure is not stable for any values of $R_1$. This figure also shows that positive and negative values of $R_1$ resulted in very similar phase diagrams.

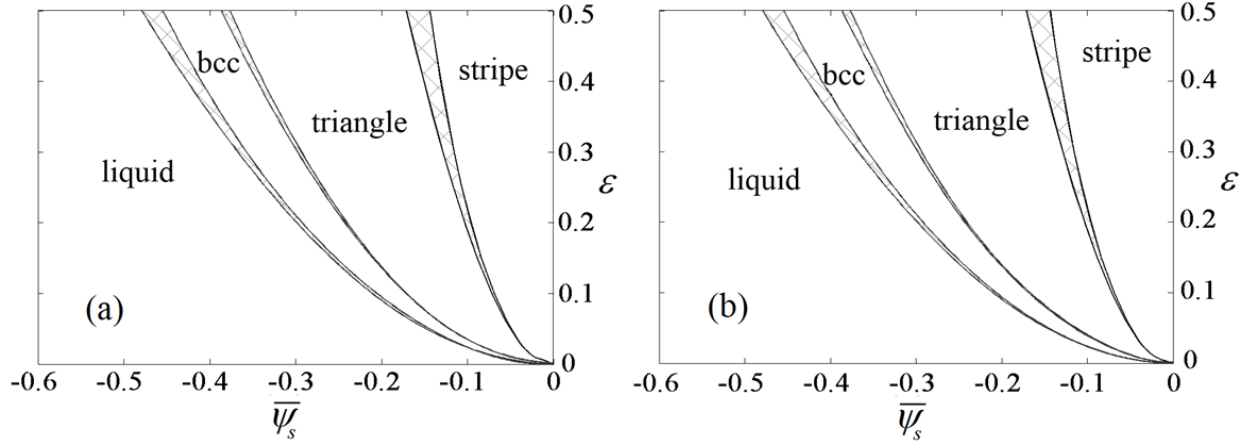

**Fig. 5.** Phase diagrams of the M2PFC model computed using two-mode expansions of the crystal density field for fcc and one-mode for bcc, triangle and stripe; $Q_1 = Q_{1,bcc} = \sqrt{2}$; (a) $R_1 = 0.05$, and (b) $R_1 = -0.015$.

## II. Phase diagram of M2PFC model considering two-mode expansion of density fields

Calculations of phase diagrams will be more accurate if the two-mode expansion of the density fields were considered for all the crystal structures, Eqs. (6a)-(6e). In this section, all the phase diagrams are calculated by two-mode expansion of density field. First, for all the crystal structures with their own $Q_1$, the phase diagrams are calculated with M2PFC model, then we recalculate the phase diagrams by considering $Q_1$ for a crystal structure as reference (first $Q_1$ for fcc and then for bcc). In M2PFC model $R_0$ will be modified for each crystal structure to exactly minimize the free energy density; therefore only accurately calculated stable phases will be presented in the phase diagrams of M2PFC model.

The value of $Q_1$ is $\sqrt{2}$, $\sqrt{3}$ and 2 for bcc, triangle, and stripe crystal structures, respectively. $R_0$ is calculated by minimizing the free energy density of the solid state for every crystal structure with respect to $q$ by considering its relevant $q$ and $Q_1$:

$$R_{0,fcc} = 2R_1(B_s^2 / A_s^2),\qquad(9a)$$

$$R_{0,bcc} = 2R_1(B_s^2 / A_s^2),\qquad(9b)$$

$$R_{0,tri} = 3R_1(B_s^2 / A_s^2),$$  (9c)

$$R_{0,str} = 4R_1(B_s^2 / A_s^2).$$  (9d)

The phase diagrams of M2PFC model for $R_1 = 0.05$ and $R_1 = -0.015$ are shown in Fig. 6. In this diagrams, $Q_1$ for each crystal structure is used. These phase diagrams show that none of the model parameters can stabilize bcc crystal structure. In these cases, the free energy density of fcc phase was always less than that of the bcc phase, primarily because $Q_1$ for fcc is less than $Q_1$ for bcc.

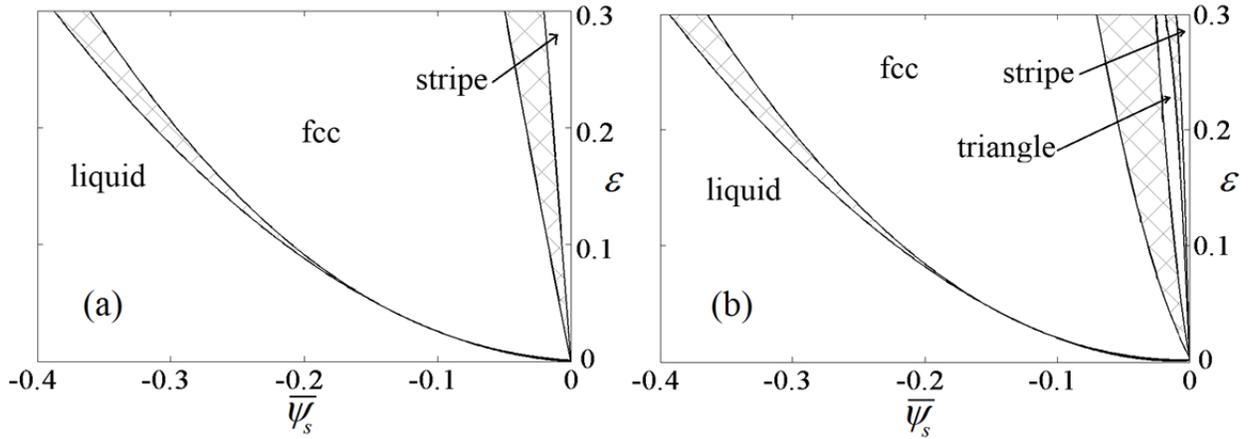

**Fig. 6.** Phase diagrams of the M2PFC model computed using two-mode expansion of density fields for all the crystal structures with their own $Q_1$; (a) $R_1 = 0.05$, and (b) $R_1 = -0.015$.

The above phase diagrams are recalculated first by considering $Q_1 = Q_{1,fcc} = \sqrt{4/3}$, and then $Q_1 = Q_{1,bcc} = \sqrt{2}$ for all the crystal structures. $R_0$ in Eq. (9a)-(9d) needs to be modified. For the case where $Q_1 = Q_{1,fcc}$, $R_0$ for fcc remains the same (Eq. (9a)), and $R_0$ for bcc, triangle and stipe is:

$$R_{0,bcc} = B_s^2(9R_1 + 10) / (3A_s^2 - 6B_s^2),$$  (10a)

$$R_{0,tri} = 2B_s^2 (9R_1 + 55)/(A_s^2 - 15B_s^2), \tag{10b}$$

$$R_{0,str} = 4B_s^2 (9R_1 + 136)/(A_s^2 - 32B_s^2). \tag{10c}$$

For the case where $Q_1 = Q_{1,bcc}$, $R_0$ for bcc remains the same (Eq. (9b)), and $R_0$ for fcc, triangle and stipe is:

$$R_{0,fcc} = B_s^2 (9R_1 + 2)/(27A_s^2 + 18B_s^2), \tag{11a}$$

$$R_{0,tri} = 6B_s^2 (R_1 + 3)/(A_s^2 - 3B_s^2), \tag{11b}$$

$$R_{0,str} = 12B_s^2 (R_1 + 10)/(A_s^2 - 8B_s^2). \tag{11c}$$

The phase diagrams for $R_1 = 0.05$ and $R_1 = -0.015$ are presented in Figs. 7 and 8. In Fig. 7(a) where $Q_1 = Q_{1,fcc}$, for $R_1 = 0.05$ and $\varepsilon < 0.33$ fcc and bcc phases are not stable, and only the triangle phase has coexistence with the liquid. By increasing the value of $\varepsilon$, the fcc phase can be stable and it has coexistence regions with liquid and triangle phases. In Fig. 7(b) with $R_1 = -0.015$, both triangle and fcc phases are stable for all the values of $\varepsilon$, but only the fcc phase has coexistence with the liquid. In Fig. 8 where $Q_1 = Q_{1,bcc} = \sqrt{2}$, for both positive and negative values of $R_1$, fcc phase is not stable. For $\varepsilon < 0.16$, Fig. 8(a) shows that only the triangle phase has a coexistence region with the liquid, and then increasing $\varepsilon$ reveals stability of the bcc phase which is in coexistence with both triangle and liquid phases. Fig. 8(b) is similar to Fig. 7(b) except instead of fcc, bcc crystal structure is stable and only fcc crystal structure has coexistence with the liquid.

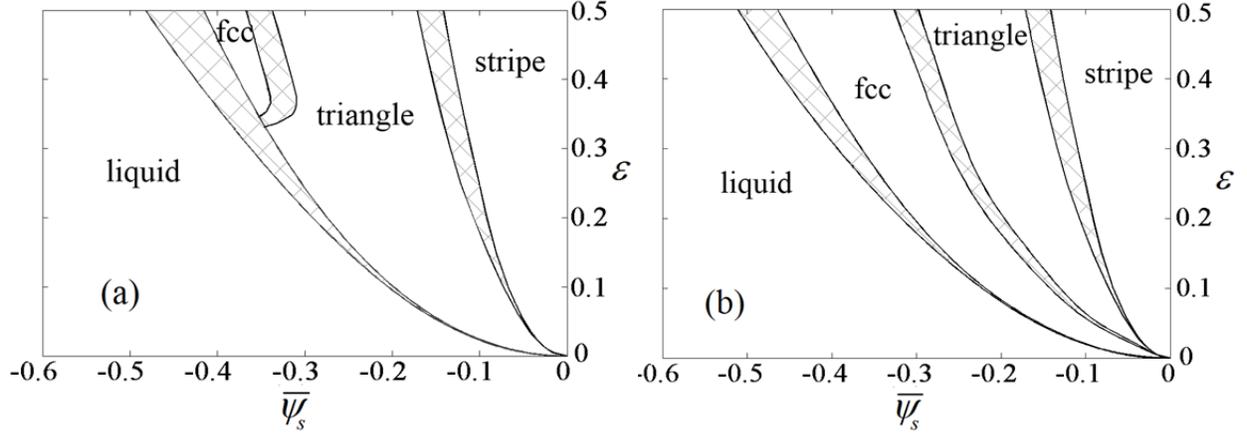

**Fig. 7.** Phase diagrams of the M2PFC model computed using two-mode expansions of density for all crystal structures and $Q_1 = Q_{1,fcc} = \sqrt{4/3}$ ; (a) $R_1 = 0.05$, and (b) $R_1 = -0.015$ .

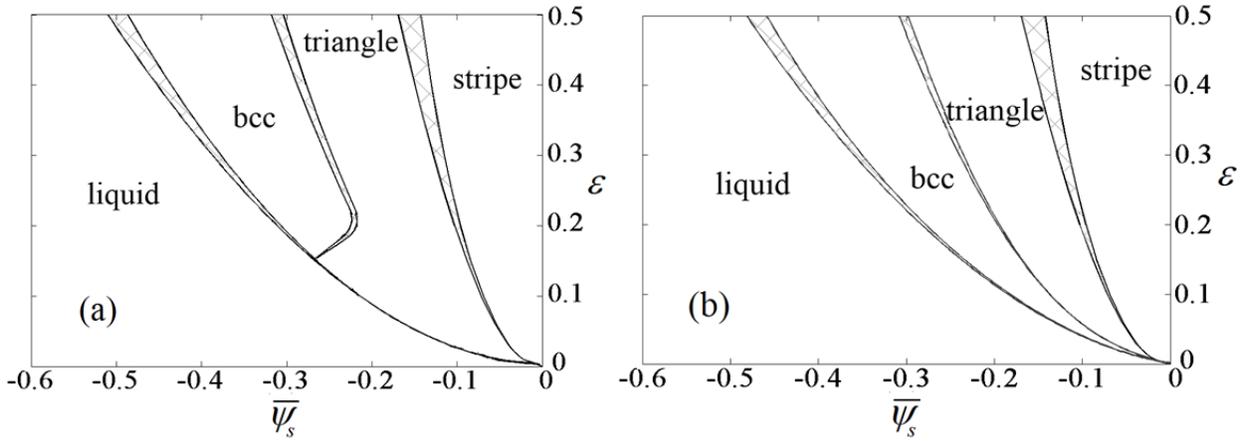

**Fig. 8.** Phase diagrams of the M2PFC model computed using two-mode expansions of density for all crystal structures and $Q_1 = Q_{1,bcc} = \sqrt{2}$ ; (a) $R_1 = 0.05$, and (b) $R_1 = -0.015$ .

### 3.2. Phase diagram in 2D

To calculate 2D phase diagrams, $Q_1$ for square is considered as the basis and the other parameters in all crystal structures are calculated. By considering $q = 1$ and $Q_1 = \sqrt{2}$ for the square structure, $R_{0,sq} = 2R_1(B_s^2 / A_s^2)$. For triangle and stripe lattice structures, $R_0$ is calculated as:

$$R_{0,tri} = 6B_s^2(R_1+3) / (A_s^2 - 3B_s^2), \tag{12a}$$

$$R_{0,str} = 12B_s^2(R_1+10) / (A_s^2 - 8B_s^2), \tag{12b}$$

$A_s$ and $B_s$ are determined by minimizing the free energy density.

For the square structure, minimizing the free energy density with respect to $A_s$ and $B_s$ leads to two coupled equations:

$$12\bar{\psi}_s^2 A_s + 48\bar{\psi}_s A_s B_s + 36A_s^3 + 72A_s B_s^2 + 4A_s R_0 R_1 + 4A_s R_0 - 4A_s \varepsilon = 0, \tag{13a}$$

$$12\bar{\psi}_s^2 B_s + 24\bar{\psi}_s A_s^2 + 36B_s^3 + 72A_s^2 B_s + 4B_s R_0 R_1 + 4B_s R_1 - 4B_s \varepsilon = 0. \tag{13b}$$

The above equations and the equation for $R_0$ of the square structure are solved simultaneously to determine $A_s$, $B_s$ and $R_0$ for different values of $R_1$, $\varepsilon$ and $\bar{\psi}_s$. This procedure was followed for the other crystal structures. The equations for triangular and stripe crystal structures are not presented here for brevity.

In Fig. 9, 2D phase diagrams of M2PFC for negative, zero and positive values of $R_1$ are plotted. Fig. 9(a) shows the constructed 2D phase diagram for $R_1 = -0.15$. In this diagram only the square phase is stable. Increasing $R_1$ to $-0.05$, decreases the square phase region but still there is no stable triangle phase ( Fig. 9(b) ). Calculated phase diagram for $R_1 = 0$ in Fig. 9(c) shows square, triangle and stripe lattice phases can be stable but no range of $\varepsilon$ allows coexistence of the triangle phase with the liquid. Calculated phase diagram in Fig. 9(d) indicates by increasing $R_1$ to a positive value of $0.05$, in addition to have both square and triangle stable phases which have a coexistence region with each other, they both have coexistence with the liquid. As mentioned before increasing $R_1$ makes the two-mode PFC behaves as one-mode PFC, and it is expected to not have a stable square phase for large values of $R_1$. Fig. 9(e) shows for $R_1 = 0.15$ only the triangle phase is stable and coexist with the liquid in $\varepsilon < 0.7$.

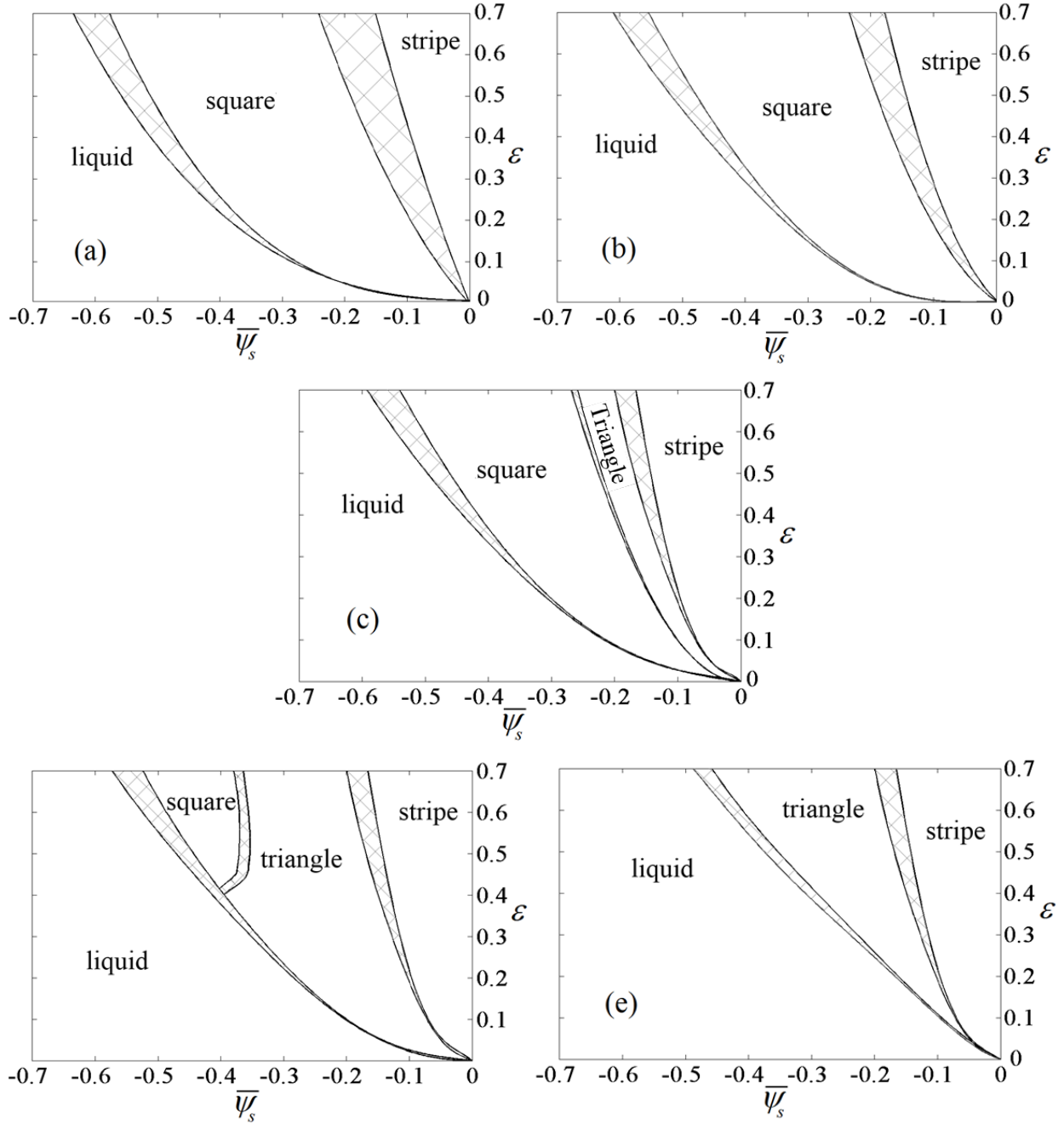

**Fig. 9.** 2D phase diagrams of the M2PFC model for (a) $R_1 = -0.15$, (b) $R_1 = -0.05$, (c) $R_1 = 0$, (d) $R_1 = 0.05$, and (e) $R_1 = 0.15$.

## 4. Conclusions

In this study phase diagrams of the modified two-mode PFC (M2PFC) model in 2D and 3D were calculated. M2PFC model by incorporating a dependent parameter $R_0$, has the ability to exactly minimize the free energy functional in each crystal structure. By presenting the phase diagrams of M2PFC model, we showed that the model is capable of simulating square, triangle, stripe, bcc and fcc lattice structures as well as their coexistence with each other and liquid phase. $R_0$ is essential in M2PFC model when two-mode expansion of density is considered. Also calculation of the phase diagram is sensitive to $Q_1$. But parameter $R_0$ gives a flexibility to the model to minimize the free energy in each crystal structure for any $Q_1$. It was shown that changing $R_1$ and $\varepsilon$ parameters allows adjustments of the relative liquid and solid free energies and densities, therefore the M2PFC model can be used to study problems related to the solid-liquid coexistence, solid state near melting point and solid state transformation.

## Acknowledgement

The authors are grateful for computer time allocation provided by the Extreme Science and Engineering Discovery Environment (XSEDE). M. Asle Zaeem would like to acknowledge the funding support from the National Science Foundation under Grant No. NSF-CMMI 1537170.

## *Appendix A. Calculation of the dimensionless density field*

Assuming that the system is in crystalline state and the average value of the density is $\bar{\psi}_s$ ,then the functional form of a periodic density can be written in terms of reciprocal lattice vectors (RLVs), $\vec{k}$ , and their amplitudes $A_{\vec{k}}$ by Eq. (5). In n-dimensional space, $\vec{k} = \sum_1^n n_i \vec{q_i}$ where $\vec{q_i}$ are the principal RLVs related to a specific crystalline symmetry and $n_i$ are integer numbers. It is appropriate to assume that the amplitudes are constant in a periodic state. A one-mode approximation will refer to an approximation in which the summation for $\vec{k}$ only includes $n_i$ that correspond to the first nearest atoms to reconstruct a given crystal symmetry and two-mode approximation includes up to the second nearest atoms. In other words, $n_i$ are chosen to include up to the $n^{th}$ order of the RLVs by considering $\sum_1^n n_i \le n$ .

In bcc crystal structure, the direct principal lattice vectors are in this form

$$\vec{a}_1 = \frac{1}{2} a (\hat{x} + \hat{y} - \hat{z}), \vec{a}_2 = \frac{1}{2} a(-\hat{x} + \hat{y} + \hat{z}), \vec{a}_3 = \frac{1}{2} a(\hat{x} - \hat{y} + \hat{z}) \; , \tag{A1}$$

Where a is lattice parameter and $\hat{x}$, $\hat{y}$ and $\hat{z}$ are the unit vectors. The relevant RLVs can be calculated by considering this relationship $a_i q_j = 2\pi \delta_{ij}$ in which $\delta_{ij}$ is the Dirac delta function.

$$\vec{q}_1 = \frac{2\pi}{a}(\hat{x} + \hat{y}), \vec{q}_2 = \frac{2\pi}{a}(\hat{y} + \hat{z}), \vec{q}_3 = \frac{2\pi}{a}(\hat{x} + \hat{z}), \tag{A2}$$

The value of $n_i$ are ones for which the magnitude of $\vec{k}$ is equal to $2\pi\sqrt{2}/a$ for one-mode approximation and $4\pi/a$ for two-mode approximation. $4\pi/a$ is equal to $\sqrt{2}$ times of the magnitude of $\vec{k}$ in one-mode. This is because bcc turns to fcc in RL space and the magnitude ratio of second RLVs to principal RLVs is $\sqrt{2}$ . So the value $n_i$ that correspond to a one-mode approximation are $(n_1, n_2, n_3) = (1,0,0), (0,1,0), (0,0,1), (1,-1,0), (0,1,-1), (-1,0,1)$. By calculating $\vec{k}$ and substituting in Eq. (5), assuming all the amplitudes are equivalent (i.e. $A_{\vec{k}} = A_s$ ) and $q = 2\pi/a$ gives:

$$\psi_{bcc} = \bar{\psi}_s + 2A_s \left[ \cos(qx + qy) + \cos(qy + qz) + \cos(qx + qz) + \cos(qx - qz) + \cos(qy - qx) + \cos(qz - qy) \right]. \tag{A3}$$

For two-mode, $(n_1, n_2, n_3) = (1,1,-1), (1,-1,1), (1,1,-1)$. With assumption of all second amplitudes are equivalent (i.e. $A_{\vec{k}} = B_s$ ), then two-mode expansion density in this case can be calculated and the dimensionless density field will be in form of Eq. (6e).

In fcc crystalline the direct principal lattice vectors, $a_i$ and RLVs have these forms,

$$\vec{a}_1 = \frac{1}{2} a (\hat{x} + \hat{y}), \vec{a}_2 = \frac{1}{2} a(\hat{y} + \hat{z}), \vec{a}_3 = \frac{1}{2} a(\hat{x} + \hat{z}) \; , \tag{A4}$$

$$\vec{q}_1 = \frac{2\pi}{a}(\hat{x} + \hat{y} - \hat{z}), \vec{q}_2 = \frac{2\pi}{a}(-\hat{x} + \hat{y} + \hat{z}), \vec{q}_3 = \frac{2\pi}{a}(\hat{x} - \hat{y} + \hat{z}) . \tag{A5}$$

The value of $n_i$ are defined in order to have the magnitude of $\vec{k}$ equals to $2\pi\sqrt{3}/a$ for one-mode approximation and $4\pi/a$ for two-mode approximation. The RLVs of fcc crystal structure has the form of bcc crystalline and the magnitude of $\vec{k}$ for second wavelength to principal is $2/\sqrt{3}$. So for one-mode approximation, $(n_1, n_2, n_3) = (1,0,0),(0,1,0),(0,0,1),(-1,-1,-1)$ and for two-mode, $(n_1, n_2, n_3) = (1,1,0),(0,1,1),(1,0,1)$. By considering previous assumption as bcc, the dimensionless density filed can be calculated in form of Eq. (A6) and simpler form of Eq. (6d).

$$\psi_{fcc} = \bar{\psi}_s + 2A_s \left[ \cos(qx + qy - qz) + \cos(-q\,x + qy + qz) + \cos(qx - qy + qz) + \cos(qx + qy + qz) \right] + 2B_s (\cos 2qx + \cos 2qy + \cos 2qz). \tag{A6}$$

For square lattice structure, $\vec{a}_1 = a\hat{x}, \vec{a}_2 = a\hat{y}$ and $\vec{q}_1 = 2\pi\hat{x}/\text{a}, \vec{q}_2 = 2\pi\hat{y}/a$. In one-mode and two-mode approximations $(n_1, n_2) = (1,0),(0,1)$ and $(n_1, n_2) = (1,1),(1,-1)$ respectively, and the dimensionless density will have this form before simplification,

$$\psi_{sq} = \bar{\psi}_s + 2A_s (\cos qx + \cos qy) + 4B_s \left[ \cos(qx + qy) + \cos(\text{q}x - qy) \right]. \tag{A7}$$

In stripe, $\vec{a}_1 = a\hat{x}, \vec{q}_1 = 2\pi\hat{x}/a$. For one-mode and two-mode expansions $n_1 = 1, -1$ and $n_1 = 2, -2$ which results the dimensionless density filed in form of Eq. (6c).

For triangle lattice structure the direct principal lattice vectors, $a_i$ and RLVs have these forms,

$$\vec{a}_1 = \frac{-2a}{\sqrt{3}}\hat{x}, \vec{a}_2 = \frac{-a}{\sqrt{3}}\hat{x} + a\hat{y}, \tag{A8}$$

$$\vec{q}_1 = \frac{-2\pi}{a}(\frac{\sqrt{3}}{2}\hat{x} + \frac{1}{2}\hat{y}), \vec{q}_2 = \frac{2\pi}{a}(\hat{y}) . \tag{A9}$$

The magnitude of $\vec{k}$ is $2\pi/a$ for one-mode approximation and $2\pi\sqrt{3}/a$ for two-mode approximation. Then $(n_1, n_2) = (1,0),(0,1),(-1,-1)$ and $(n_1, n_2) = (-1,-2),(-2,-1),(-1,1)$ are for one and two mode expansion approximations. If $q = 2\sqrt{3}\pi / 2a$, these set of vectors leads to the following approximation for density and the simpler form of Eq. (6b).

$$
\begin{aligned}
\psi_{tri} = \bar{\psi}_s + A_s &\left[\cos(qx + \frac{qy}{\sqrt{3}}) - \cos\frac{2qy}{\sqrt{3}} + \cos(qx - \frac{qy}{\sqrt{3}})\right] \\
&+ B_s\left[\cos(qx - \sqrt{3}qy) + \cos(qx + \sqrt{3}qy) - \frac{1}{2}\cos 2qx\right].
\end{aligned} \tag{A10}
$$

In our dimensionless units the magnitude of the principal RLVs are unity, so for square and stripe lattice structures, $q = 1$, for the triangular, $q = \sqrt{3} / 2$, for bcc, $q = 1/\sqrt{2}$ and for fcc, $q = 1/\sqrt{3}$.

***Appendix B.*** *The free energy density in solid state*

By substituting $\psi$ from Eqs. (6a)-(6e) in $F^*$ in Eq. (8b), integrating over the crystal structure, and calculate the free energy per unit area in 2D or per unit volume in 3D, the free energy in crystalline state can be calculated. The expression of $f_s$ in square, triangle, stripe, bcc and fcc lattice structures are defined as shown below. For the liquid free energy, $F_l$, it is simpler. By considering a constant $\psi_l$ in Eq. (8b), the dimensionless liquid free energy can be calculated.

$$
\begin{aligned}
F_{bcc} = \frac{\bar{\psi}_s^4}{4} &- \left[\varepsilon - (1 + R_0)(Q_1^4 + R_1)\right]\frac{\bar{\psi}_s^2}{2} + 6\left[(1 + Q_1^4 - 2Q_1^2)R_0 + R_0R_1 - \varepsilon + 3\bar{\psi}_s^2\right]A_s^2 \\
&+ 3\left[(4 + Q_1^4 - 4Q_1^2)R_0 + R_0R_1 - \varepsilon + 3\bar{\psi}_s^2 - 4Q_1^2 + Q_1^4 + R_1 + 4\right]B_s^2 \\
&+ 72A_s^2B_s\bar{\psi}_s + 144A_s^2B_s^2 + 144A_s^3B_s + 48A_s^3\bar{\psi}_s + 135A_s^4 + \frac{45}{2}B_s^4,
\end{aligned} \tag{B1}
$$

$$
\begin{aligned}
F_{fcc} = \frac{\bar{\psi}_s^4}{4} &- \left[\varepsilon - (1 + R_0)(Q_1^4 + R_1)\right]\frac{\bar{\psi}_s^2}{2} + 4\left[(1 + Q_1^4 - 2Q_1^2)R_0 + R_0R_1 - \varepsilon + 3\bar{\psi}_s^2\right]A_s^2 \\
&+ 3\left[(\frac{16}{9} + Q_1^4 - \frac{8}{3}Q_1^2)R_0 + R_0R_1 - \varepsilon + 3\bar{\psi}_s^2 - \frac{8}{81}Q_1^2 + \frac{1}{9}Q_1^4 + \frac{1}{9}R_1 + \frac{16}{81}\right]B_s^2 \\
&+ 72A_s^2B_s\bar{\psi}_s + 144A_s^2B_s^2 + 54A_s^4 + \frac{45}{2}B_s^4,
\end{aligned} \tag{B2}
$$

$$F_{str} = \frac{\overline{\psi}_s^4}{4} - \left[ \varepsilon - (1+R_0)(Q_1^4 + R_1) \right] \frac{\overline{\psi}_s^2}{2} + \left[ (1+Q_1^4 - 2Q_1^2)R_0 + R_0 R_1 - \varepsilon + 3\overline{\psi}_s^2 \right] \frac{A_s^2}{4}$$

$$+ \left[ (16 + Q_1^4 - 8Q_1^2)R_0 + R_0 R_1 - \varepsilon + 3\overline{\psi}_s^2 + 144 - 72Q_1^2 + 9Q_1^4 \right] \frac{B_s^2}{4} \tag{B3}$$

$$+ \frac{3}{4} A_s^2 B_s \overline{\psi}_s + \frac{3}{8} A_s^2 B_s^2 + \frac{3}{32} A_s^4 + \frac{3}{32} B_s^4 ,$$

$$F_{tri} = \frac{\overline{\psi}_s^4}{4} - \left[ \varepsilon - (1+R_0)(Q_1^4 + R_1) \right] \frac{\overline{\psi}_s^2}{2} + \left[ (1+Q_1^4 - 2Q_1^2)R_0 + R_0 R_1 - \varepsilon + 3\overline{\psi}_s^2 \right] \frac{3}{16} A_s^2$$

$$+ \left[ (9 + Q_1^4 - 6Q_1^2)R_0 + R_0 R_1 - \varepsilon + 3\overline{\psi}_s^2 - 24Q_1^2 + 4Q_1^4 + 4R_1 + 36 \right] \frac{3}{16} B_s^2 \tag{B4}$$

$$- \frac{3}{16} A_s^3 \overline{\psi}_s - \frac{3}{16} B_s^3 \overline{\psi}_s + \frac{9}{64} A_s^3 B_s + \frac{45}{128} A_s^2 B_s^2 + \frac{45}{512} A_s^4 + \frac{45}{512} B_s^4 ,$$

$$F_{sq} = \frac{\overline{\psi}_s^4}{4} - \left[ \varepsilon - (1+R_0)(Q_1^4 + R_1) \right] \frac{\overline{\psi}_s^2}{2} + \left[ (1+Q_1^4 - 2Q_1^2)R_0 + R_0 R_1 - \varepsilon + 3\overline{\psi}_s^2 \right] 2A_s^2$$

$$+ \left[ (4 + Q_1^4 - 4Q_1^2)R_0 + R_0 R_1 - \varepsilon + 3\overline{\psi}_s^2 - 4Q_1^2 + Q_1^4 + R_1 + 4 \right] 2B_s^2 \tag{B5}$$

$$+ 24 A_s^2 B_s \overline{\psi}_s + 36 A_s^2 B_s^2 + 9 A_s^4 + 9 B_s^4 ,$$

$$F_l = - \left[ \varepsilon - (1+R_0)(Q_1^4 + R_1) \right] \frac{\psi_l^2}{2} + \frac{\psi_l^4}{4} . \tag{B6}$$